# Extension of Wald-Wolfowitz Runs Test for Regression Validity Testing with Repeated Measures of Independent Variable

Bo-Yao Lian[a] and Nelson G. Chen[a,b]*

[a]*Institute of Intelligent Bioelectrical Engineering, National Yang Ming Chiao Tung University, Hsinchu, Taiwan;* [b]*Department of Electronics and Electrical Engineering, National Yang Ming Chiao Tung University, Hsinchu, Taiwan*

CONTACT Nelson G. Chen ✉ *ngchen@nycu.edu.tw* 🏢 National Yang Ming Chiao Tung University, Hsinchu, Taiwan

# Extension of Wald-Wolfowitz Runs Test for Regression Validity Testing with Repeated Measures of Independent Variable

The Wald-Wolfowitz runs test can assess the correctness of a regression curve fitted to a data set with one independent parameter. The assessment is performed through examination of the residuals, where the signs of the residuals would appear randomly if the regression curve were correct. We propose extending the test to the case where multiple data points were measured for specific independent parameter values. By randomly permutating the data points corresponding to each independent parameter value and treating their residuals as occurring in their permutated sequence and then executing the runs test, results are shown to be equivalent to those of a data set containing the same number of points with no repeated measurements. This approach avoids the loss of points, and hence loss of test sensitivity, were the means at each independent parameter value used. It also avoids the problem of weighting each mean differently if the number of data points measured at each parameter value is not identical.

Keywords: runs test; regression; residuals; goodness-of-fit

**1. Introduction**

Given a data set with a single independent variable (e.g., time t) and a dependent variable y that is measured for various values of the independent variable, one commonly performs a regression procedure on the data to obtain an equation y = f(t) that describes the responses of y as a function of the independent variable. The equation can be linear or nonlinear, depending on the underlying phenomenon being modelled. Regardless, such an equation will not fit the data points "perfectly," where $f(t_i)$ exactly equals the measured corresponding $y_i$. Instead, errors will be collected in a residual term $\varepsilon_i = y_i - f(t_i)$ for every point. Traditionally, the curve-fitting process obtains constant parameters for the underlying equation, minimizing the sum of squares of the residual values.

Additional usual assumptions are (1) the independent variable values were measured perfectly, (2) the residuals ε have a mean value of zero and have a constant variance (homoscedasticity), and (3) each value of $\varepsilon_i$ is independent of the other $\varepsilon_i$ values (Poole and O'Farrell 1971). If the residuals are not independent of each other, then there would be a pattern to their appearance. A lack of randomness indicates a poor fit, and this lack of randomness can be assessed through the Wald-Wolfowitz runs test (1940) as applied to a binary sequence (Swed and Eisenhart 1943). Numerous texts (Bard 1974; Draper and Smith 1998; Cobelli, Foster, and Toffolo 2002; Cleophas and Zwinderman 2013) have described testing the randomness of the sequence of the signs of the residuals as a means of determining the adequacy of a given regression. Examples of this approach being used include its application toward medical longitudinal data (Chang 2000), near-infrared spectroscopy (Centner, de Noord, and Massart 1998), and dry matter degradation in animals (Uckardes, Korkmaz, and Ocal 2013). It is also part of the Graphpad Prism software package used to assess the adequacy of a linear or nonlinear curve fit.

However, the runs test as currently applied to a curve fit requires that the measured data points have unique values of the independent variable (e.g. time t). (For the remainder of this paper, the independent variable will be referred to as time for simplicity, even though the analysis remains valid for other independent variables as well.) In other words, repeated measurements cannot be at any given timepoint. While one could, in theory, apply the runs test toward the calculated means at each point, doing so leads to two problems.

(1) Suppose the number of measured points at certain time points are different. In that case, the calculated means will be of varying quality (heteroskedastic) depending on the number of points used to compute each average. Treating these means as if they were homoscedastic would therefore be incorrect.

(2) The number of values examined for runs is reduced, leading to the test not fully

utilizing the fact that there were the measured number of independent points.

The runs test applied to regression, the issue with repeated measurements at a given time point, and the two problems are illustrated in Figure 1.

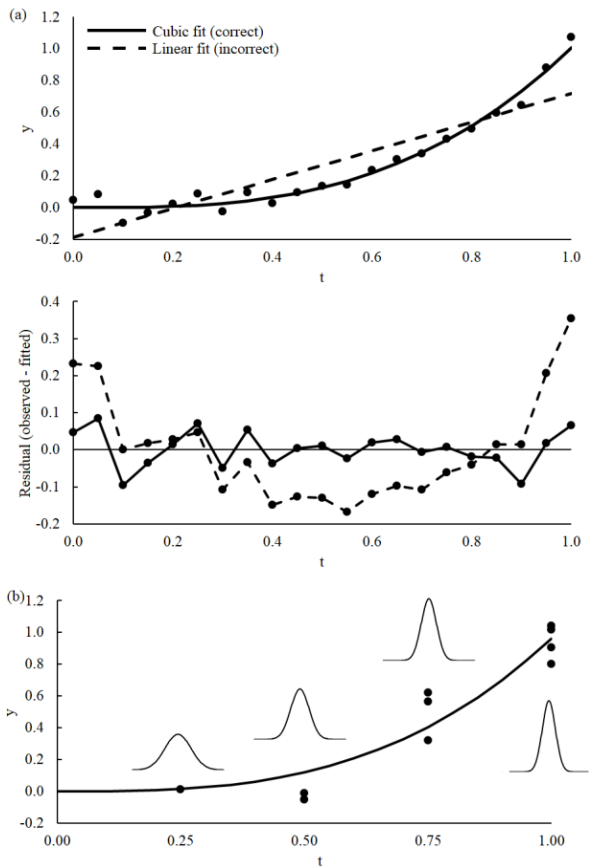

**Figure 1.** Panel (a) depicts the runs test as applied to assessing two regression models. The underlying relation depicted is $y(t) = t^3$ with the correct fitted curve of the form $At^3$ shown as a solid line, and an incorrect linear curve of the form $At + B$ shown as a dashed line. The residuals from the cubic model form 13 runs about zero ($p = 0.5779$), while the linear model forms only 3 runs ($p = 1.19 \times 10^{-4}$), which indicates its incorrectness. Note how there

is only a single measurement at any timepoint. Panel (b) illustrates a situation where data is obtained repeatedly at multiple timepoints, with 1 to 4 measurements depending on the time. The resulting distribution of averages taken at each timepoint differs, due to the change in standard errors going from timepoint to timepoint. Also, if the means at each timepoint (0.25, 0.50, 0.75, and 1.00) were used to assess the fitted curve using the runs test, there would be only 4 total means present for an assessment when there are in reality 10 independent points that can be used.

## 2. Random Permutation of Data Point Sequence

To address the problems described above, we propose randomly permutating the "sequence" in which the data points appear, whenever they are measured at the same timepoint. A schematic of the concept is provided in Figure 2.

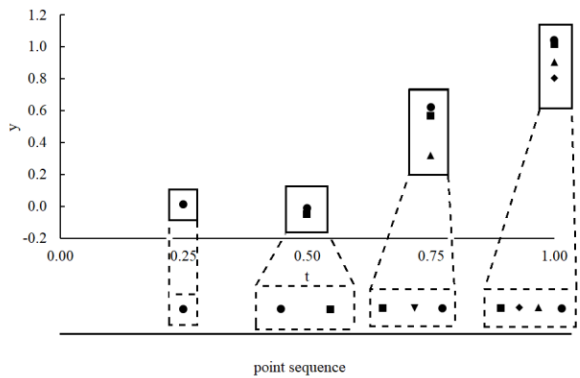

**Figure 2.** Shown is a schematic of random permutation as applied to repeatedly measured points. Each measured point is plotted with a different symbol for any specified time. If only one datapoint was measured at a particular timepoint, such as t = 0.25, then no changes are made. Otherwise, the sequence that the points appear at each specific timepoint is treated as a random permutation of all the measurements made. The figure shows 2, 3,

and 4 measured points at times 0.5, 0.75, and 1, respectively. A randomly permutated sequence used for analysis is plotted at the bottom.

After the points are treated thus, the quality of the regression is assessed using the runs test in the usual manner, looking at the number of runs of the residuals.

This treatment of the repeated measures at any timepoint t can be justified by noting that each measured point has a corresponding actual time $t + \varepsilon_t$, with $\varepsilon_t$ being an unknown small quantity. Since the values of $\varepsilon_t$ are unknown, the underlying order of the points is equally likely to be any given permutation of these points. If the measured data are all above or below the regression curve, permutating the points would not affect the number of runs present. If the measured data span the regression curve, the number of runs could increase depending on the specific random permutation selected; however, the number of increases is not expected to be large relative to the total number of runs.

## 3. Computer Simulation

Computer simulations were performed to examine the validity of this proposed extension of the runs test. All simulations used the arbitrarily selected underlying function $f(t) = 2t + 1$ as the basis for randomly generated values. Normally distributed simulated noise with a mean of zero and standard deviation of 0.34 was added to each simulated data point. Simulations were performed using MATLAB (The Mathworks, Inc., Natick, MA, USA).

### *3.1. Constant Number of Repeated Measurements per Timepoint*

Random data sets were generated from the underlying function for 2, 3, 4, and 5

repeated measurements at each timepoint, with 3-14 total timepoints (5-14 for cases with 2 repeated measurements). (Each data point had the value $f(t) = 2t + 1 + \varepsilon_{random}$.) Timepoints t were 1, 2, 3, …, 14 with the maximum t equal to the total number of timepoints. For instance, for 10 timepoints, timepoints t are 1, 2, 3, …, 10. The data was then fit using least-squares fitting to the linear expression $f(t) = At + B$, with A and B being constants to be estimated. Next, the resulting residuals were computed. For each given timepoint, residual values were randomly permutated to obtain a sequence. The number of runs of these residuals was recorded. There were no cases of any residual being exactly equal to zero. This process was then repeated so that for each case (e.g. 3 repeated measurements at each timepoint with 6 total timepoints) there were 100,000 trials. Cases were selected to have at least 9 data points, since 9 points are the minimum number needed to demonstrate significance at the $\alpha = 0.05$ level with two runs.

The distribution of runs for each case was next examined. Approximate 95% and 99% confidence intervals were estimated as follows. Starting with their mode, the smallest whole number that, when added and subtracted from the mode, would cover more than 95% and 99% of the observed runs, was noted and these ranges of runs were recorded. An example is provided in Table 1.

**Table 1.** Calculation of approximate 95% and 99% confidence intervals (CI) for the number of runs for the case of 4 measurements at each timepoint and 4 timepoints. Adding the number of cases where there were 4-14 runs (inclusive) gives 99,796 cases, the smallest number of cases centered around the mode that exceeds 99% of all cases.

| Runs | Number (out of 100,000) | Notes |
| --- | --- | --- |
| 3 | 78 | |
| 4 | 338 | lower-bound (inclusive) for 99% CI |
| 5 | 1894 | lower-bound (inclusive) for 95% CI |
| 6 | 4955 | |
| 7 | 11818 | |
| 8 | 17094 | |
| 9 | 21683 | Mode |
| 10 | 18426 | |
| 11 | 13527 | |
| 12 | 6642 | |
| 13 | 2776 | upper-bound (inclusive) for 95% CI |
| 14 | 643 | upper-bound (inclusive) for 99% CI |
| 15 | 117 | |
| 16 | 9 | |

A second set of simulations was performed with the same underlying function f(t) = 2t + 1 corresponding to each case previously described, with the same total number of data points but with no repeated timepoints. Data points were evenly spaced between 1 and the maximum time of the corresponding case. Just as before, least-squares best-fit lines were computed and the number of the runs of the residuals noted. For each case, the simulation was run 100,000 times. For example, the corresponding simulation for the 4 repeated measurements at 4 timepoints case above has a total of 16 points evenly spaced in time

between 1 and 4 (1, 1.2, 1.4, …, 3.8, 4.0).

The distribution of the runs of the residuals with no repeated timepoints was recorded and compared with the distribution of the runs of the residuals of the corresponding case using the two-sample runs test (Wald and Wolfowitz 1940) with tie values (same number of runs in both distributions) randomly permutated. Computed p-values were recorded to determine whether the distributions were statistically different, with the Šidák correction applied to the critical p-value to account for multiple comparisons.

*3.2. Differing Number of Repeated Measurements per Time Point*

Four additional cases were run with different numbers of repeated measurements per timepoint to determine whether the previous results can be generalized to these cases. The cases had various numbers of measurements at each timepoint, and these numbers are provided in Table 2.

**Table 2.** Test cases with an unequal number of datapoints at various timepoints.

| Case/Timepoints | 1 | 2 | 3 | 4 | 5 | 6 | 7 | 8 | 9 | 10 | 11 | 12 | 13 | 14 |
|---|---|---|---|---|---|---|---|---|---|---|---|---|---|---|
| 1 | | 2 | 3 | 3 | 3 | 3 | 3 | 3 | 3 | | | | | |
| 2 | | 2 | 4 | 5 | 3 | 5 | 3 | 4 | 5 | 5 | 5 | 5 | 4 | 5 | 3 |
| 3 | | 5 | 5 | 5 | 4 | 5 | 4 | 5 | 4 | 5 | 3 | 5 | 4 | 5 | 3 |
| 4 | | 4 | 5 | 5 | 5 | 4 | 5 | 4 | 5 | 5 | 5 | 5 | 5 | 5 | 5 |

As before, simulated data was generated for the underlying expression f(t) = 2t + 1, with the same randomly distributed error term ε added to each datapoint. The simulated data were fitted using least-squares fitting to the linear expression f(t) = At + B, and residuals were computed with the residuals corresponding to each timepoint randomly permuted. The distribution of the number of runs was recorded and compared to the distribution of runs for the same number of evenly spaced datapoints with no repeated measurements. The simulation was also run 100,000 times for each case.

### *3.3. Sufficiency of Running Each Simulation 100,000 Times*

To verify that running each simulation 100,000 times provides sufficiently accurate distribution information, where the confidence intervals for the number of runs of the residuals do not change with additional simulation runs, five cases were selected for examination. Case 1 had 11 timepoints with 2 datapoints per timepoint. Case 2 was case 1 from Table 2, with differing numbers of datapoints per timepoint. Case 3 had 9 timepoints with 5 datapoints per timepoint. Case 4 had 12 timepoints with 4 datapoints per timepoint. Finally, case 5 had 10 timepoints with 5 datapoints per timepoint. Each case was rerun 1,000,000 times, and the distribution of runs was noted. The approximate 95% and 99% confidence intervals of the numbers of runs were computed for these cases and compared with those from corresponding cases run 100,000 times.

## 4. Simulation Results

### *4.1. Constant Number of Repeated Measurements per Timepoint*

Approximate 95% and 99% confidence intervals of the number of runs and the p-values of the comparison of the distribution of runs versus the case with no repeated timepoints are given in Table 3.

**Table 3.** Confidence intervals of the distribution of runs and comparison to cases without repeated timepoints. The following information is presented in each cell from top to bottom: approximate 95% confidence interval, 99% confidence interval (both inclusive), and p-value of the comparison to an equivalent case without repeated timepoints. P-values less than 0.0011/0.0002 would indicate familywise statistically significant differences at the $\alpha = 0.05/0.01$ levels after application of the Šidák correction.

| number of timepoints/points per timepoint | 2 | 3 | 4 | 5 |
|---|---|---|---|---|
| 3 |  | N/A[a] | 4-10 | 5-13 |
|  |  |  | 3-11 | 5-13 |
|  |  | 0.2448 | 0.0306 | 0.0034 |
| 4 |  | 4-10 | 5-13 | 7-15 |
|  |  | 3-11 | 4-14 | 6-16 |
|  |  | 0.1032 | 0.0798 | 0.4264 |
| 5 | 3-9 | 5-13 | 7-15 | 8-18 |
|  | 3-9 | 5-13 | 6-16 | 7-19 |
|  | 0.1886 | 0.2215 | 0.0294 | 0.9155 |
| 6 | 4-10 | 6-14 | 8-18 | 11-21 |
|  | 3-11 | 5-15 | 7-19 | 9-23 |
|  | 0.0818 | 0.5027 | 0.7552 | 0.1361 |
| 7 | 5-13 | 7-15 | 10-20 | 13-25 |
|  | 4-14 | 5-17 | 9-21 | 12-26 |
|  | 0.5205 | 0.9134 | 0.1826 | 0.3314 |
| 8 | 5-13 | 5-18 | 12-22 | 15-27 |
|  | 4-14 | 4-19 | 10-24 | 13-29 |
|  | 0.1971 | 0.2519 | 0.0564 | 0.4617 |
| 9 | 5-13 | 10-20 | 13-25 | 16-30 |
|  | 3-15 | 9-21 | 12-26 | 14-32 |
|  | 0.2059 | 0.6406 | 0.4159 | 0.5560 |
| 10 | 7-15 | 12-22 | 15-27 | 20-34 |
|  | 6-16 | 10-24 | 13-29 | 18-36 |
|  | 0.3544 | 0.8833 | 0.9584 | 0.3916 |
| 11 | 8-18[b] | 11-23 | 17-29 | 22-36 |
| | 7-19 | 10-24 | 15-31 | 20-38 |
| | 0.6490 | 0.5312 | 0.2283 | 0.3445 |
| 12 | 8-18 | 13-25 | 18-32 | 24-38 |
| | 7-19 | 12-26 | 16-34 | 21-41 |
| | 0.8508 | 0.5116 | 0.4211 | 0.9568 |
| 13 | 10-20 | 15-27 | 20-34 | 25-41 |
| | 8-22 | 13-29 | 18-36 | 23-43 |
| | 0.8401 | 0.1503 | 0.6084 | 0.7915 |
| 14 | 10-20 | 17-29 | 22-36 | 29-45 |
| | 9-21 | 15-31 | 20-38 | 26-48 |
| | 0.4920 | 0.1294 | 0.5489 | 0.7495 |

[a]This case did not produce approximate 95% or 99% confidence intervals. The total number of runs ranged from 3-9; no subrange centered around the mode would encompass these percentages of values.

[b]This case initially produced a p-value of 0.0008, with approximate confidence intervals as shown. Repeating the simulation 1,000,000 times, rather than the original 100,000, resulted in the p-value given in the table. Approximate confidence intervals were unchanged.

No statistically significant differences were found between the distribution of runs with and without repeated timepoints.

*4.2. Differing Number of Repeated Measurements per Timepoint*

Results of the four selected cases, as indicated in Table 2, are provided in Table 4.

**Table 4.** Confidence intervals of distribution of runs and comparison to cases without repeated timepoints of four cases with differing numbers of repeated measurements per time point. P-values less than 0.0127/0.0025 would indicate familywise statistically significant differences at the $\alpha = 0.05/0.01$ levels after application of the Šidák correction.

| Case | 1 | 2 | 3 | 4 |
|---|---|---|---|---|
| approximate 95% confidence interval | 9-17 | 24-38 | 24-40 | 27-43 |
| approximate 99% confidence interval | 7-19 | 21-41 | 22-42 | 25-45 |
| p-value | 0.2505 | 0.1149 | 0.8041 | 0.1257 |

As with the cases with a constant number of measurements per timepoint, there were no detectable differences between these cases and corresponding cases with no repeated timepoints.

*4.3. Sufficiency of Running Each Simulation 100,000 Times*

The approximate confidence intervals of each case were unchanged upon running each simulation 100,000 times versus running them 1,000,000 times, as indicated in Table 5. This lack of difference demonstrates the sufficiency of running each simulation 100,000 times.

**Table 5.** Approximate confidence intervals (CIs) of the number of runs (all inclusive).

| Case | 1 | 2 | 3 | 4 | 5 |
| --- | --- | --- | --- | --- | --- |
| Number of points | 22 | 23 | 45 | 48 | 50 |
| 95% CI | 8-18 | 9-17 | 16-30 | 18-32 | 20-34 |
| 99% CI | 7-19 | 7-19 | 14-32 | 16-34 | 18-36 |
| (100,000 runs) | | | | | |
| 95% CI | 8-18 | 9-17 | 16-30 | 18-32 | 20-34 |
| 99% CI | 7-19 | 7-19 | 14-32 | 16-34 | 18-36 |
| (1,000,000 runs) | | | | | |

**5. Discussion and Conclusion**

We demonstrated that one could treat repeated measurements at each value of the independent variable as values in a sequence to be randomly permuted to assess regression validity. We did so by showing that the distribution of the number of runs for various cases with repeated measurements does not differ from cases with the same number of all non-repeated measurements. This extension to the Wald-Wolfowitz runs test avoids problems with averaging the data at each timepoint, where the number of data points would be effectively reduced and may introduce problems caused by the different distribution of averages at different locations.

Our study examined a single underlying function from which datasets were simulated. However, the results may be generalized to other underlying functions, as long as the standard assumptions (independence of points, homoscedasticity, residual mean of zero) of regression analysis remain valid. Correctly fitting a different underlying function will produce residuals that behave the same. The number of runs would not be too small or large for valid regressions, as assessed through the runs test. Conversely, an incorrect regression would produce too few or too many runs in the signs of the residuals.

The random permutation of the repeated measurements as described has each possible permutation at a given timepoint have equal probability of being selected. In practice, it is recommended that the permutation(s) used for the runs test be selected by machine or some other random method to avoid bias.